\documentclass[aps,prl,reprint,superscriptaddress,amsmath,floatfix]{revtex4-1}
\usepackage{circuitikz}
\usepackage[utf8]{inputenc}
\usepackage{amsmath,amssymb}
\usepackage{bm}
\usepackage{siunitx}
\usepackage{ifpdf}
\usepackage{graphicx}
\usepackage{float}
\usepackage{amsthm}
\usepackage{braket}
\usepackage{mathtools}
\usepackage[driverfallback=dvipdfmx,hypertexnames=false,bookmarks=false,colorlinks=true,linkcolor=blue,citecolor=blue]{hyperref} %for local
\newcommand{\average}[1]{\ensuremath{\langle#1\rangle} }

\newcommand\numberthis{\addtocounter{equation}{1}\tag{\theequation}}
\usepackage{version}	% required for `\comment' (yatex added)
\begin{document}
\title{Theory of nonreciprocal Josephson effect}

\author{Kou Misaki}
\affiliation{Department of Applied Physics, The University of Tokyo, 
Bunkyo, Tokyo 113-8656, Japan}
\author{Naoto Nagaosa}
\affiliation{Department of Applied Physics, The University of Tokyo, 
Bunkyo, Tokyo 113-8656, Japan}
\affiliation{RIKEN Center for Emergent Matter Science (CEMS), Wako, Saitama 351-0198, Japan}
\date{\today}
\begin{abstract}
The hetero junction between the different materials often gives the rectification effect; e.g., pn junction is used for diode. On the other hand, the Josephson junction between the two superconductors is assumed to show symmetric response between the two directions of the current, i.e., the voltage drop $V$ is anti-symmetric with respect to the sign change of the current $I$. However, there should be an asymmetry between the states of charge accumulation on right and left sides of the Josephson junction, which can lead to the nonreciprocal responses. Here we demonstrate theoretically that nonreciprocal $I$-$V$ characteristic appears due to this charging energy difference both in the classical and quantum regimes. This result will pave a route to design and develop the Josephson diode.
\end{abstract}

\maketitle
\section{Introduction}
The nonreciprocal responses in noncentrosymmetric materials in general have been intensively studied both from the theoretical and the experimental viewpoint \cite{tokura_nonreciprocal_2018}. It often happens that broken $\cal{T}$ is needed in addition to broken $\cal{P}$ to obtain the nonreciprocal responses, but there are cases where only $\cal{P}$ breaking is enough. The pn junction is a representative example, where the hetero junction of n-type and p-type semiconductors acts as a rectifier without magnetic field or magnetization. On the other hand, the direction of the arrow of time is determined by the dissipation associated with the resistivity, i.e., irreversibility. In the case of  pn-junction, the existence of the depletion layer due to the Coulomb interaction is essential for its rectification function. Another example of the nonreciprocal response without $\cal{T}$-breaking is the Zener tunneling \cite{kitamura_nonreciprocal_2019}. In this case, the inter-band tunneling probability across the band gap differs between right and left directions due to the shift vector originating from the Berry connection \cite{berry_quantal_1984} even without the broken $\cal{T}$. This shift vector is also relevant to the shift current for the interband photoexcitation \cite{morimoto_topological_2016}. Therefore, the quantum geometry, which encodes the information of the microscopic inversion asymmetry inside a unit cell, plays an important role. The nonreciprocity in optical systems has been widely studied \cite{jalas_what_2013}, and in particular, the quantum diode of light has been theoretically studied \cite{fratini_fabry-perot_2014} and experimentally realized \cite{rosario_hamann_nonreciprocity_2018}. Here the two isolated two level system act as nonlinear mirrors and lead to left-right asymmetric 
Fabry-Perot interferometer. Also, as for the Josephson junctions, there are studies on the Josephson diode \cite{carapella_ratchet_2001,majer_quantum_2003,Raissi_josephson_diode,Recher_josephson_diode,Hassler_josephson_diode,Dai_josephson_diode,savelev_experimentally_2002,martinez-perez_efficient_2013}, but the nonreciprocity of voltage drop of the single Josephson junction has not been studied.

Josephson effect is a representative macroscopic quantum phenomenon where the superconducting current depends on the phase difference $\varphi$ of the order parameter between the two superconductors. The dynamics of $\varphi$ in the dissipationless case is described by the following Hamiltonian:
\begin{equation}
    H= \frac{Q^2}{2C} + E_J\left(1-\cos\frac{2e\phi}{\hbar}\right)-I_{x}\phi, \label{eq:Josephson_Hamil}
\end{equation}
where $\phi=\hbar\varphi/(2e)$, $C$ is the capacitance of the Josephson circuit,
$Q$ is the charge accumulated at the capacitance, $E_J$ is the Josephson coupling energy, $-e < 0$ is the charge of an electron, $[\phi,Q]=i\hbar$, and $I_x$ is the external current bias, which is assumed to be constant. Here we assumed the symmetric charging energy $Q^2/(2C)$, i.e., $Q$ and $-Q$ are equivalent. We will discuss the consequences of the asymmetric charging energy later. Eq. \eqref{eq:Josephson_Hamil} can be regarded as the Hamiltonian of a particle under the tilted cosine type potential with the period $\delta\phi=\pi\hbar/e$, where $Q$ and $\phi$ represent the momentum and the position, respectively. When $I_x$ is small, near the local minimum, the potential energy can be approximated by the one of the harmonic oscillator where the mass $m=C$ and the characteristic frequency $\omega=(2e/\hbar)\sqrt{E_J/C}$. Then, the width of the wavefunction is given by $\Delta \phi = \sqrt{\hbar/(m\omega)}$, and the overlap of the wavefunction between the adjacent minima is negligible when $\Delta \phi \ll \delta\phi\Leftrightarrow E_J/E_Q \gg 1$ (case (I)), and large when $\Delta \phi \gg \delta\phi\Leftrightarrow E_J/E_Q \ll 1$ (case (II)), where $E_Q=e^2/(2C)$. We also include the resistive shunt, and the Josephson circuit we will discuss is schematically shown in Fig. \ref{fig:josephson_circuit} in the Supplementary Materials.

In the case (I), $\phi$ is well-localized inside the minima, and including the resistive shunt, the dynamics is described by the semiclassical Josephson equation given by \cite{tinkham_introduction_2004}
\begin{equation}
    \hbar \dot{\varphi} =  2eV,
\label{eq:varphidot}
\end{equation}
\begin{equation}
     \dot{Q} |_{\rm cap.} +  I_c\sin\varphi + \frac{V}{R} = I_x, 
\label{eq:qdot}
\end{equation}
where $I_c=2e E_J/\hbar$, $Q|_{\rm cap.}$ is the charge accumulated at the capacitance, $V$ is the chemical potential (voltage) drop, and $R$ is the shunt resistance. Here we neglected the quantum decay probability, which is known \cite{caldeira_quantum_1983} to be expressed as $P\propto\exp[-A E_J/(\hbar \omega) ]=\exp[-A \sqrt{E_J/(8E_Q)} ]$ at zero temperature in the dissipationless case, where $A$ is the constant factor. We note that the dissipation further suppresses the quantum decay probability \cite{caldeira_quantum_1983}.

\begin{figure}
    \centering
    \includegraphics[width=\columnwidth]{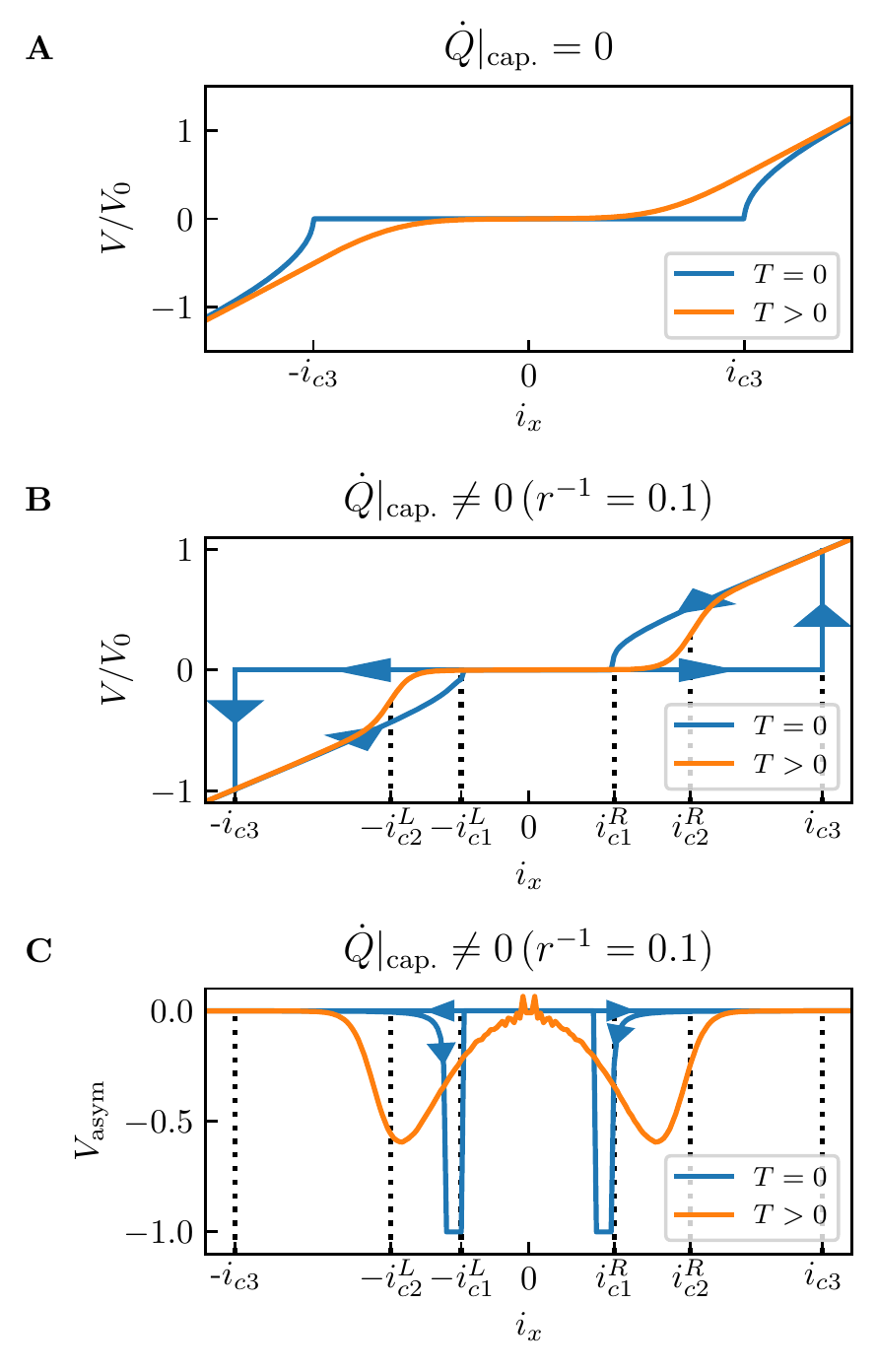}
    \caption{\textbf{$I_x-V$ curve for case (I) at $T=0$ and $T>0$.} $I_x-V$ curve calculated by the classical Langevin equation \eqref{rsjeq2} and \eqref{asymchaene} (case (I)), for the system (\textbf{A}) without $\dot{Q}|_{\rm cap.}$ term and (\textbf{B} and \textbf{C}) with $\dot{Q}|_{\rm cap.}$ term (for the definition of $r^{-1}$, see Eq. \eqref{rsjeq2}), where $i_x = I_x/I_c$ and $V_0=R I_c$ with $I_c=2eE_J/\hbar$. In (\textbf{C}), we show $V_{\rm asym}(i_x) = [V(i_x)+V(-i_x)]/[V(i_x)-V(-i_x)]$ which quantifies the degree of nonreciprocity calculated from the $I_x-V$ curve (\textbf{B}). We note that $V_{\rm asym}=0$ identically for the $I_x-V$ curve (\textbf{A}), i.e., when $\dot{Q}|_{\rm cap.}=0$. The arrows on blue curves represents the direction of the sweep of $i_x$. We set $\tilde{T} = 0.25$ for $T>0$ data, i.e., for orange curves, and $A=0.6$, $A'=0.3$ in Eqs.  \eqref{rsjeq2} and \eqref{asymchaene}.}
    \label{schematic}
\end{figure}

In the absence of the capacitance, i.e.,  $\dot{Q}|_{\rm cap.}=0$, the $I_x-V$ characteristic is solved easily to be $V=0$ for $|I_x|<I_c= 2eE_J/\hbar$ and the time-averaged voltage $\bar{V}={\rm sign}(I_x) R\sqrt{I_x^2 - I_c^2 }$ for $|I_x|>I_c= 2eE_J/\hbar$. Therefore, there occurs no nonreciprocal response in this case. In the presence of the capacitance $C$, i.e., $Q|_{\rm cap.}= CV$, the differential equation becomes second order, i.e., the inertia term appears, which results in the coexistence of the two solutions for a range of $I_x$ and hysteresis behavior of $I_x-V$ characteristic, see Fig. \ref{schematic}B, blue curve. We will numerically show that, in this case, the nonreciprocal $I_x-V$ curve is realized if we include the effect of the asymmetry coming from $\dot{Q}|_{\rm cap.}$. To understand why $\dot{Q}|_{\rm cap.}$ term is necessary for the nonreciprocal effect, here we discuss the inversion symmetry, $\mathcal{P}$, and the time reversal symmetry, $\mathcal{T}$, of Eqs. \eqref{eq:varphidot} and \eqref{eq:qdot}, in the absence of $\dot{Q}|_{\rm cap.}$ term. $\cal{T}$ transforms $I_x \to - I_x$, $\varphi \to  -\varphi$, while $V \to V$ as we can see from Eq. \eqref{eq:varphidot}. Note here that the last term on the l.h.s. of Eq. \eqref{eq:qdot} changes sign when $\cal{T}$ is applied, although $V$ is even with respect to $\cal{T}$. This is usual since $1/R$ represents the dissipation and irreversiblity, and introduces the asymmetry between the two directions of time. As for the inversion symmetry $\cal{P}$, on the other hand, the transformation gives $I_x \to - I_x$, $\varphi \to - \varphi$, and $V \to - V$ since the two superconductors are exchanged. Therefore, the nonreciprocal response, it it exists, comes from the term $\dot{Q} |_{\rm cap.}$ in Eq. \eqref{eq:qdot} when the spatial inversion symmetry $\cal{P}$ is broken. 

In the case (II), since the cosine potential is small, $Q$ is almost the good quantum number. In the same spirit as the nearly free electron approximation, $E_J(1-\cos[2e\phi/\hbar])$ term in the Hamiltonian can be treated perturbatively, and it leads to the Bragg reflection and opens up a gap at the momentum $Q=\pm \hbar\pi/\delta \phi=\pm e$. The size of the gap is proportional to $E_J$, and the energy at Brillouin zone edge is $E_Q$, so the dimensionless quantity $E_J/E_Q$ is roughly the ratio of the bandgap to the bandwidth. The last term in Eq. \eqref{eq:Josephson_Hamil} can be regarded as the potential coming from the external electric field $E=I_x$, and, including the dissipation term, the dynamics is described by
\begin{equation}
    \frac{dQ}{d t} = I_x - \frac{1}{R}\frac{\partial \tilde{E}_{ch}(Q)}{\partial Q}, \label{eq:dqdteq}
\end{equation}
where $\tilde{E}_{ch}(Q)$ is the band energy with the gap at $Q=\pm e$.

In the present paper, we study theoretically  the nonreciprocal nature of $I_x$-$V$ characteristics of the asymmetric Josephson junction, which is modeled by the asymmetric charging energy  $E_{ch}(Q)(\ne E_{ch}(-Q))$. We will show that, both for case (I) and case (II), the asymmetry of $E_{ch}(Q)$ leads to the nonreciprocity.

Before getting into the detailed analysis, here we discuss the origin of the asymmetric charging energy. The capacitance of small junction system originates from two contributions: One is the classical capacitance, determined by the electrostatic energy inside the thin film, and the other one is the quantum capacitance, which depends on the property of the charge response of two sandwiching bulk systems \cite{buttiker_mesoscopic_1993,buttiker_capacitance_1993,christen_low_1996,ma_weakly_1999,wang_nonlinear_1999}. Among these two contributions, the latter one is in general nonlinear. In the discussion section, we will estimate the order of the quantum capacitance in real systems and discuss how to experimentally measure the nonreciprocity discussed in the main text.

\section{Results}
\subsection{Nonreciprocal $I_x-V$ curve at $T=0$ for case (I)}
In Fig. \ref{schematic} (blue curves), we show the $I_x-V$ curve for the system without the $\dot{Q} |_{\rm cap.}$ term (panel A) and the system with $\mathcal{P}$ breaking $\dot{Q} |_{\rm cap.}$ term (panel B) at $T = 0$. As we mentioned in the introduction, the nonreciprocity is realized only for the latter system, see panel C.

An important feature of $I_x-V$ curve at $T=0$ with finite $\dot{Q}|_{\rm cap.}$ (Fig. \ref{schematic}B, blue curve) is the hysteresis at $i_{c1}^R<i_x<i_{c3}$ and $-i_{c3}^L<i_x<-i_{c1}^L$. This comes from the coexistence of the limit cycle and the stable fixed point \cite{hanggi_reaction-rate_1990,strogatz_nonlinear_2018}. As can be seen from Figs. \ref{fig:bifs_schematic_underdamped}B and C in the Supplementary Materials, because of the presence of the limit cycle, for the initial condition inside the dark blue region, the stationary state at long time is governed by the limit cycle so that the finite voltage drop results. On the contrary, for the initial condition inside the green region, the particle is attracted to the fixed point and the voltage drop is zero. Sweeping $i_x$ from the large value to the small value corresponds to the former case, while sweeping $i_x$ from the small value to the large value corresponds to the latter case. Namely, the hysteresis behavior occurs.
On the contrary, there exists no hysteresis for $I_x-V$ curve at $T=0$ without $\dot{Q}|_{\rm cap.}$ term (Fig. \ref{schematic}A, blue curve). 

Here we review the qualitative aspect of the bifurcation of limit cycle in the system with $T=0$ \cite{hanggi_reaction-rate_1990,strogatz_nonlinear_2018} for $i_x > 0$. The system shows qualitatively different behavior depending on the value of the dissipation strength $r$, defined above Eq. \eqref{rsjeq2}.

For $r^{-1} \gg 1$ (Fig. \ref{schematic}A), we can neglect the inertia term (the capacitance) and the equation becomes
\begin{equation}
 r^{-1}d\varphi/d\tau=i_x-\sin\varphi. \label{eq:overdamped}
\end{equation}
For $i_x>1$, $d\varphi/d\tau>0$ and there is only a limit cycle (Fig. \ref{fig:bifs_schematic_overdamped}C in the Supplementary Materials). At $i_x=i_{c3} = 1$, the saddle-node (blue-sky) bifurcation leads to the vanishing of the limit cycle and the birth of the stable and unstable fixed points at $\varphi=\sin^{-1}i_x$ and $\pi-\sin^{-1}i_x$ for $i_x<1$, respectively, see Figs. \ref{fig:bifs_schematic_overdamped} B and C in the Supplementary Materials. For $i_x<1$, the long time dynamics is governed by the stable fixed point, see Figs. \ref{fig:bifs_schematic_overdamped}A and B in the Supplementary Materials. Therefore, in this case the disappearance of the limit cycle and the birth of the stable fixed point occur simultaneously, i.e., $i_{c1}^R=i_{c3}=1$. Above $i_{c3}$, the flow of $\varphi$ occurs, and the time-average of $d \varphi/d \tau$ gives that of the voltage drop $\bar{V}={\rm sign}(I_x) R\sqrt{I_x^2 - I_c^2 }$ as we mentioned in the introduction.

For $r^{-1}\ll 1$ (Fig. \ref{schematic}B), we cannot neglect the inertia term (the capacitance) and the bifurcation mentioned above splits into two bifurcations. One is at $i_x=i_{c3} = 1$, where the saddle-node bifurcation leads to the birth of the stable fixed point and the saddle point at $(\varphi,q)=(\sin^{-1}i_x,0)$ and $(\pi-\sin^{-1}i_x,0)$, as is shown in Figs. \ref{fig:bifs_schematic_underdamped} C, D and E in the Supplementary Materials; The other one is the homoclinic bifurcation at $i_x=i_{c1}^R$, where the limit cycle collides with the saddle point at $(\varphi,q)=(\pi-\sin^{-1}i_x,0)$ to become the homoclinic orbit and then disappears, as is shown in Figs. \ref{fig:bifs_schematic_underdamped}A and B in the Supplementary Materials. We will review what a homoclinic orbit is and discuss its role in the phase diagram later. As for the bifurcations for $i_x < 0$, the qualitative nature of the bifurcations are the same, but importantly, $i_{c1}^L \neq i_{c1}^R$ because of the asymmetry of the charging energy. It leads to the enhancement of $V_{\rm asym}$ near $i_{c1}^L$ and $i_{c1}^R$ as can be seen in Fig. \ref{schematic}C.

\begin{figure*}
    \centering
    \includegraphics[width=0.85\textwidth]{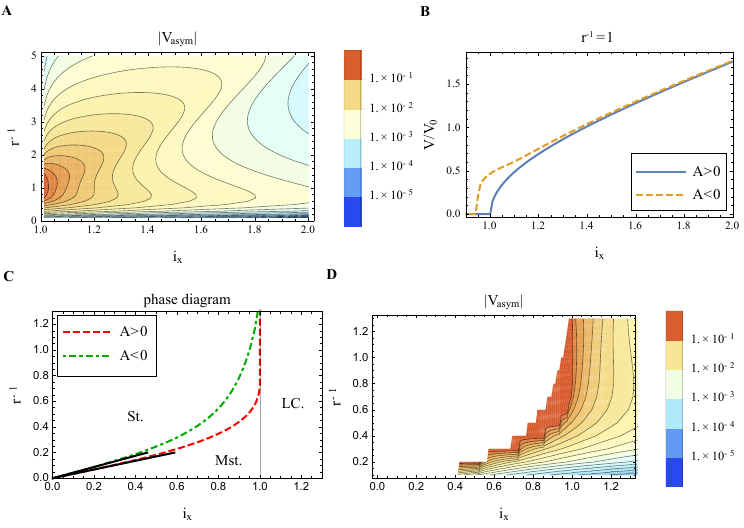}
    \caption{\textbf{Nonreciprocity for various $i_x$ and $r^{-1}$ for case (I) at $T=0$.} (\textbf{A}) $V_{\rm asym}(i_x) = [V(i_x)+V(-i_x)]/[V(i_x)-V(-i_x)]$ as a function of $i_x$ and $r^{-1}$ calculated by Eqs. \eqref{rsjeq2} and \eqref{asymchaene} with $\tilde{T} = 0$. (\textbf{B}) The voltage drop $V/V_0$ where $V_0=RI_c$ for $A>0$ and $A<0$ with $r^{-1}=1$ and $\tilde{T} = 0$ in Eqs. \eqref{rsjeq2} and \eqref{asymchaene}. Here $V_{\rm asym} < 0$ for the parameter region shown in (\textbf{A}). (\textbf{C}) The phase diagram in $(i_x,r^{-1})$ space for Eqs. \eqref{rsjeq2} and \eqref{asymchaene} with $\tilde{T} = 0$. St., Mst. and LC. represent the phase with stable fixed point only, stable fixed point coexisting with limit cycle, and limit cycle only, respectively. The black curves are the phase boundary calculated from Eq. \eqref{melnikov}. (\textbf{D}) $V_{\rm asym}(i_x) = [V(i_x)+V(-i_x)]/[V(i_x)-V(-i_x)]$ near the phase boundary, where $V(i_x)$ is calculated for the metastable limit cycle of  Eqs. \eqref{rsjeq2} and \eqref{asymchaene} with $\tilde{T} = 0$, i.e., the plot corresponds to the sweeping of $i_x$ from the large value in Fig. \ref{schematic}D. $V_{\rm asym}<0$ for the parameter region shown in (\textbf{B}).}
    \label{vasdata}
\end{figure*}
\subsection{Nonreciprocity for various $i_x$ and $r^{-1}$ at $T=0$ for case (I)}
For $|i_x|>1$, $V_{\rm asym}$ as a function of $i_x$ and $r^{-1}$ is shown in Fig. \ref{vasdata}A. We can see that the nonreciprocity is enhanced for small $i_x$ and $r^{-1}$. Since $|i_x|>1$, the dynamics is governed by the limit cycle traversing from $\varphi=-\pi$ to $\pi$ at finite $q$ as is shown in Figs. \ref{stablemfd}A and B in the Supplementary Materials. As we can see, finite $A$ modifies the limit cycle and leads to the asymmetry. 

For $|i_x|<1$, the homoclinic bifurcation occurs at $i_{c1}^R$ and $-i_{c1}^L$. As we explained in the last section, at this bifurcation point the limit cycle becomes the homoclinic orbit. In short, a homoclinic orbit is a variant of a limit cycle. However, in contrast to a limit cycle, there exists a fixed point on it, so its time period is infinite, since it takes infinite time to reach and depart from the fixed point. For example, the black curves in Figs. \ref{fig:bifs_schematic_underdamped}B and \ref{homcli}A and B in the Supplementary Materials are homoclinic orbits where the fixed point is shown by red dots. In our case, the presence of the homoclinic orbit indicates the homoclinic bifurcation, so by identifying the parameter where there exists a homoclinic orbit on the $(i_x,r^{-1})$ plane, we can identify the phase boundary. 

For small $i_x$ and $r^{-1}$, we can perturbatively calculate the phase boundary from the parameter point $i_x=r^{-1}=0$, where we can analytically obtain the homoclinic orbit, see Figs. \ref{homcli}A and B in the Supplementary Materials. For that, we calculate the simple zero of the following Melnikov function \cite{guckenheimer_nonlinear_1983}:
\begin{align}
    &\int_{-\infty}^{\infty}dt \dot{\varphi_0}(t)(i_x-r^{-1}\dot{\varphi_0}(t))\nonumber\\
    =& 2\pi i_x - 2r^{-1} \int_0^{q_{\rm max}}dq 
\left(\frac{d\epsilon_{ch}(q)}{dq}\right)^2
\frac{1}{\sqrt{\epsilon_{ch}(q)[2-\epsilon_{ch}(q)]}}, \numberthis \label{melnikov}
\end{align}
where $\varphi_0(t)$ is the homoclinic orbit for $i_x=r^{-1}=0$ shown in Fig. \ref{homcli} in the Supplementary Materials, and $q_{\rm max}$ is the maximum of $q$ along the orbit. As we can see, the homoclinic orbit for $A>0$ (Fig. \ref{homcli}A, black curve) and $A<0$ (Fig. \ref{homcli}B, black curve) is quite different and that leads to the difference of the Melnikov function and the phase boundary in two cases. In Fig. \ref{vasdata}C, we show the phase boundary obtained from direct numerical calculation (red dotted and green dot-dashed curves) and the one obtained from the condition that Eq. \eqref{melnikov} should be zero (black solid curve). We can see that the prediction of Eq. \eqref{melnikov} agrees well with the numerically obtained boundary for small $i_x$ and $r^{-1}$. For $(i_x,r^{-1})$ such that metastable limit cycle does exist for $A<0$ but not for $A>0$, we observe very large $|V_{\rm asym}|$, as is shown in Fig. \ref{vasdata}D, since the time-averaged velocity $\overline{d\varphi/d\tau}=0$ for $A>0$ but $\overline{d\varphi/d\tau}$ is finite for $A<0$. We also note that the large $|V_{\rm asym}|$ for $i_x \gtrsim 1$ (Fig. \ref{vasdata}A) can be understood as a consequence of the difference of $i_{c1}$ for $A>0$ and $A<0$: As we can see from Fig. \ref{vasdata}B, the voltage drop $V$ is larger at $i_x \gtrsim 1$ for $A<0$, because $i_{c1}$ is smaller for $A<0$.

\subsection{Nonreciprocal $I_x-V$ curve at finite temperature $T>0$ for case (I)}
For the finite temperature $T>0$ case, we numerically simulated the Langiven equation Eq. \eqref{rsjeq} with stochastic Heun's scheme \cite{garcia-palacios_langevin-dynamics_1998} to calculate the physical quantities and then took an ensemble average. Numerically calculated $I_x-V$ curve is shown in Fig. \ref{schematic} (orange curves). As is shown in Fig. \ref{schematic}B, we can see that the voltage drop $V$ suddenly increases around $i_{c2}^R$ and $-i_{c2}^L$ and merges to the curve $V/V_0=i_x$. This behavior can be understood as the dynamical transition, from the state where the dominant probabilistic weight is on the stable fixed point so that the voltage drop is around zero, to the one where the limit cycle is primarily realized and the finite voltage drop results \cite{risken_fokker-planck_1996,hanggi_artificial_2009}. Since the system is at the finite temperature, the transition is not sharp, but as $T\to +0$ this transition becomes sharper and sharper and the jump of $V$ from $0$ to finite value occurs at $i_x=i_{c2}^{R}$ and $-i_{c2}^L$ when $T=+0$. At the same time, the relaxation time between the two configurations diverges as $T\to +0$, and when the experimental measurement time is smaller than the relaxation time, we observe the hysteresis behavior as we discussed above for $T=0$ case. In the similar manner to $T=0$ case, the large $V_{\rm asym}$ near $i_{c2}^R$ and $-i_{c2}^L$ is realized because $i_{c2}^R\neq i_{c2}^L$.

\subsection{Nonreciprocity for various $i_x$ and $r^{-1}$ at $T >0$ for case (I)}
\begin{figure}
    \centering
    \includegraphics[width=\columnwidth]{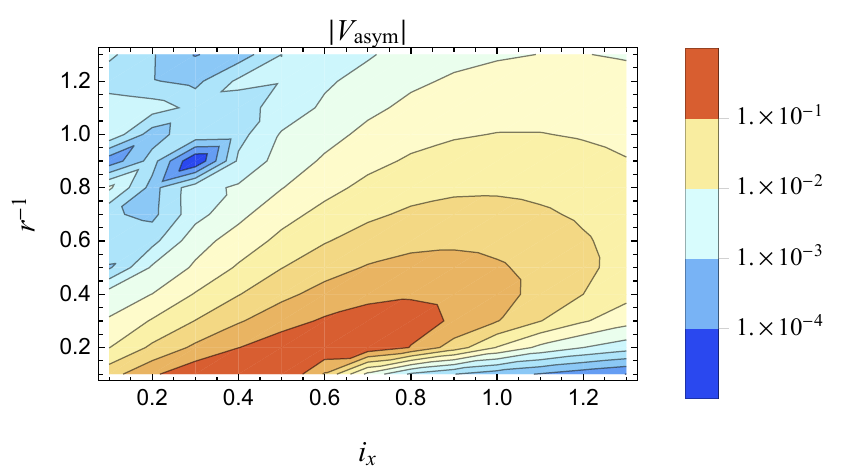}
    \caption{\textbf{Nonreciprocity for various $i_x$ and $r^{-1}$ for case (I) at $T>0$.} $V_{\rm asym}(i_x) = [V(i_x)+V(-i_x)]/[V(i_x)-V(-i_x)]$ at finite temperature. For the parameter region shown in the plot, $V_{\rm asym} < 0$. We used Eqs. \eqref{rsjeq2} and \eqref{asymchaene} with $\tilde{T} = 0.25$.}
    \label{finitetempfig}
\end{figure}

We numerically calculated the nonreciprocity for various $i_x$ and $r^{-1}$, and the result of the numerical calculation is shown in Fig. \ref{finitetempfig}.
\begin{figure*}
    \centering
    \includegraphics{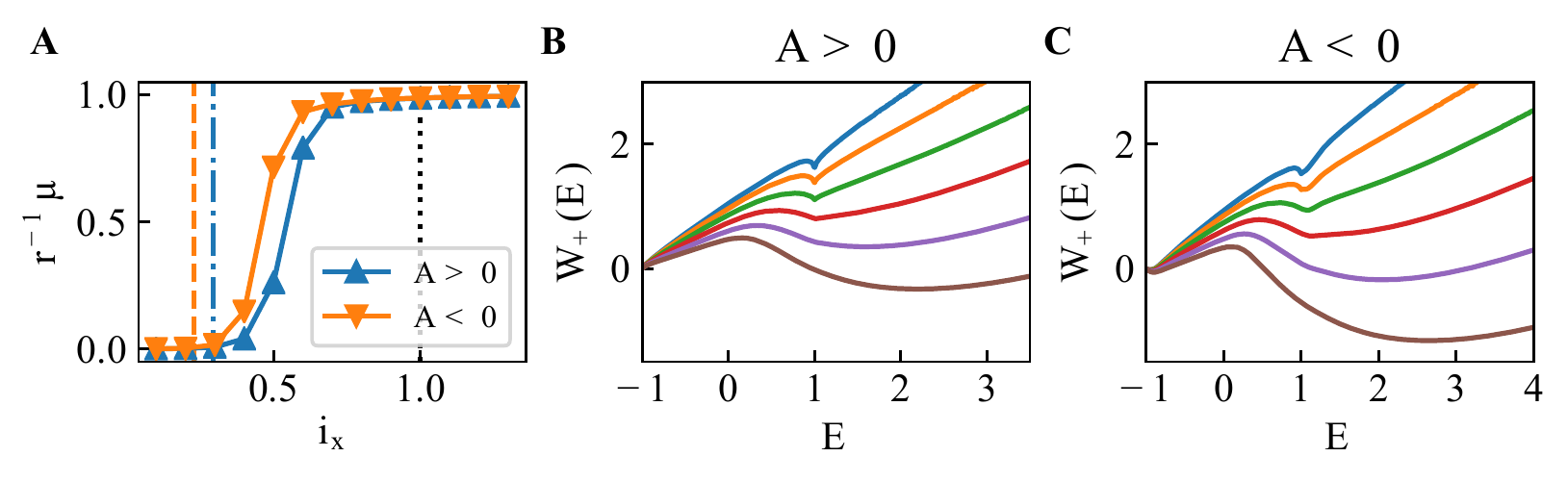}
    \caption{\textbf{Normalized mobility and large deviation function for case (I) at $T>0$.} (\textbf{A}) Normalized mobility $r^{-1}\mu$, where $r^{-1}\mu=V/(V_0 i_x)$ and $V_0=RI_c$, as a function of $i_x$ for fixed $r^{-1}$. The value where the limit cycle appears is shown by the dot-dashed blue and the dashed orange curves, and the value where the stable fixed point vanishes is shown by black dotted curve. 
    (\textbf{B} and \textbf{C}) $W_+(E)$ from $i_x=0.1$ (blue curve) to $i_x=0.6$ (brown curve), for $A>0$ and $A<0$. The parameters are set to be $r^{-1}=0.1$, $\tilde{T} = 0.25$ in Eqs. \eqref{rsjeq2} and \eqref{asymchaene}.}
    \label{potshapefig}
\end{figure*}

As we can see, the nonreciprocity is enhanced for small $r^{-1}$, i.e., small dissipation, region. This is consistent with the fact that, for $r^{-1}\gg 1$, we can neglect the inertia term in Eq. \eqref{rsjeq2} to obtain the usual inversion-symmetric overdamped Langevin equation. In addition, we can see the peak structure at finite value of $i_x$ for fixed $r^{-1}$. To understand this behavior, it is useful to plot the normalized mobility $r^{-1}\mu = V/(V_0 i_x)$, where $V_0=RI_c$, as a function of $i_x$ \cite{risken_fokker-planck_1996}, see Fig. \ref{potshapefig}A. We can see that for small $i_x$, the mobility is almost zero, but at some finite $i_x$ the mobility jumps to $\mu = r$ and saturates. This kind of behavior can be understood from the large deviation function of the energy $W_{\pm}(E)$, defined as
\begin{align}
    P(E) = \left\{\begin{array}{c}
        \mathcal{N}_+ e^{-W_+(E)/\tilde{T}}\quad (q \geq 0)  \\
        \mathcal{N}_- e^{-W_-(E)/\tilde{T}}\quad (q < 0)
    \end{array}
    \right.,\,
    E=\epsilon_{ch}(q) - \cos\varphi.
\end{align}
where $\mathcal{N}_{+/-}$ is the normalization factor, $P(E)$ is the distribution function of $E$, and we introduced two functions $W_+$ and $W_-$, corresponding to the two branches of momentum $q$ as a function of the energy $E$ \cite{risken_fokker-planck_1996}. Numerically calculated $W_+(E)$ for $A>0$ and $A<0$ is shown in Figs. \ref{potshapefig}B and C. We can see that, as we increase the bias $i_x$, $W_+(E)$ at large $E$ becomes small and eventually the local minimum at $E>1$ drops below the value at $E=-1$. This corresponds to the dynamical transition of the typical trajectory from the static one at $E=-1$ to the running one at $E>1$. We can see that the critical value of $i_x$ which we denote $i_{c2}$, where this transition occurs is different for $A>0$ case ($i_{c2}\sim 0.6$) and $A<0$ case ($i_{c2} \sim 0.5$). The fact that $i_{c2}$ is larger for $A>0$ is consistent with the larger $i_{c1}$ where the limit cycle emerges, as is shown by blue dot-dashed and orange dashed curves in Fig. \ref{potshapefig}A.

\begin{figure}
    \centering
    \includegraphics{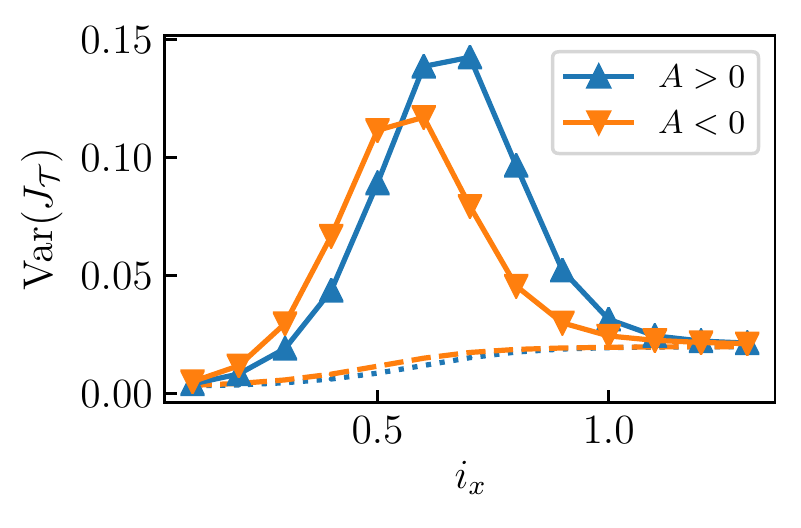}
    \caption{\textbf{Variance of $J_\mathcal{T}$ for case (I) at $T>0$.} Variance of $J_\mathcal{T}$ as a function of $i_x$. The blue dotted and orange dashed curves are the lower bound predicted by the thermodynamic uncertainty relation, ${\rm Var}(J_{\mathcal{T}}) \geq 2\average{J_{\mathcal{T}}}^2 /(\mathcal{T} \sigma) $, where $\sigma$ is the entropy production rate and is calculated as $\sigma=i_x \average{J_{\mathcal{T}}}/\tilde{T}$ \cite{fischer_large_2018}. We numerically simulated the Langevin equation \eqref{rsjeq2} and \eqref{asymchaene} for 100 ensembles with time $\tau = 10^7$ and $\Delta \tau = 10^{-2}$ by the stochastic Heun scheme, and set $\mathcal{T} = 1000$. The parameters are set to be $r^{-1}=0.1$, $\tilde{T} = 1$.}
    \label{jtfig}
\end{figure}
Because of the presence of the thermal fluctuation, we can discuss not only the average value of the velocity, but also the whole distribution of the time-averaged current $J_{\mathcal{T}}=\int_0^{\mathcal{T}}d\tau \frac{d\varphi}{d\tau}$ \cite{fischer_large_2018}. The numerically calculated variance is shown in Fig. \ref{jtfig}. Since the system does not have $\mathcal{T}'$ symmetry (For the definition of $\mathcal{T}'$ and $\mathcal{P}'$ symmetry, see Materials and Methods.), we might have a violation of the lower bound of the variance known as thermodynamic uncertainty relation \cite{barato_thermodynamic_2015,pietzonka_universal_2016,fischer_large_2018}, as is observed in the underdamped Langevin system with magnetic field \cite{chun_effect_2019}, but we did not observe any violation as far as for the parameter regions we have checked. As we can see, the fluctuation of the current becomes large for intermediate $i_x$. This reflects the fact that there coexists the stationary trajectory and the running trajectory, and these two trajectories, which have quite different average velocities, are probabilistically realized, leading to the large fluctuation of the current. For larger $i_x$ the fluctuation decreases, since the stationary fixed point disappears. Reflecting the difference of the critical current $i_{c2}$, the region where the current fluctuation enhances is different for $A>0$ and $A<0$ cases, and that leads to quite different current fluctuation as we can see in Fig. \ref{jtfig}.

\begin{figure*}
    \centering
    \includegraphics{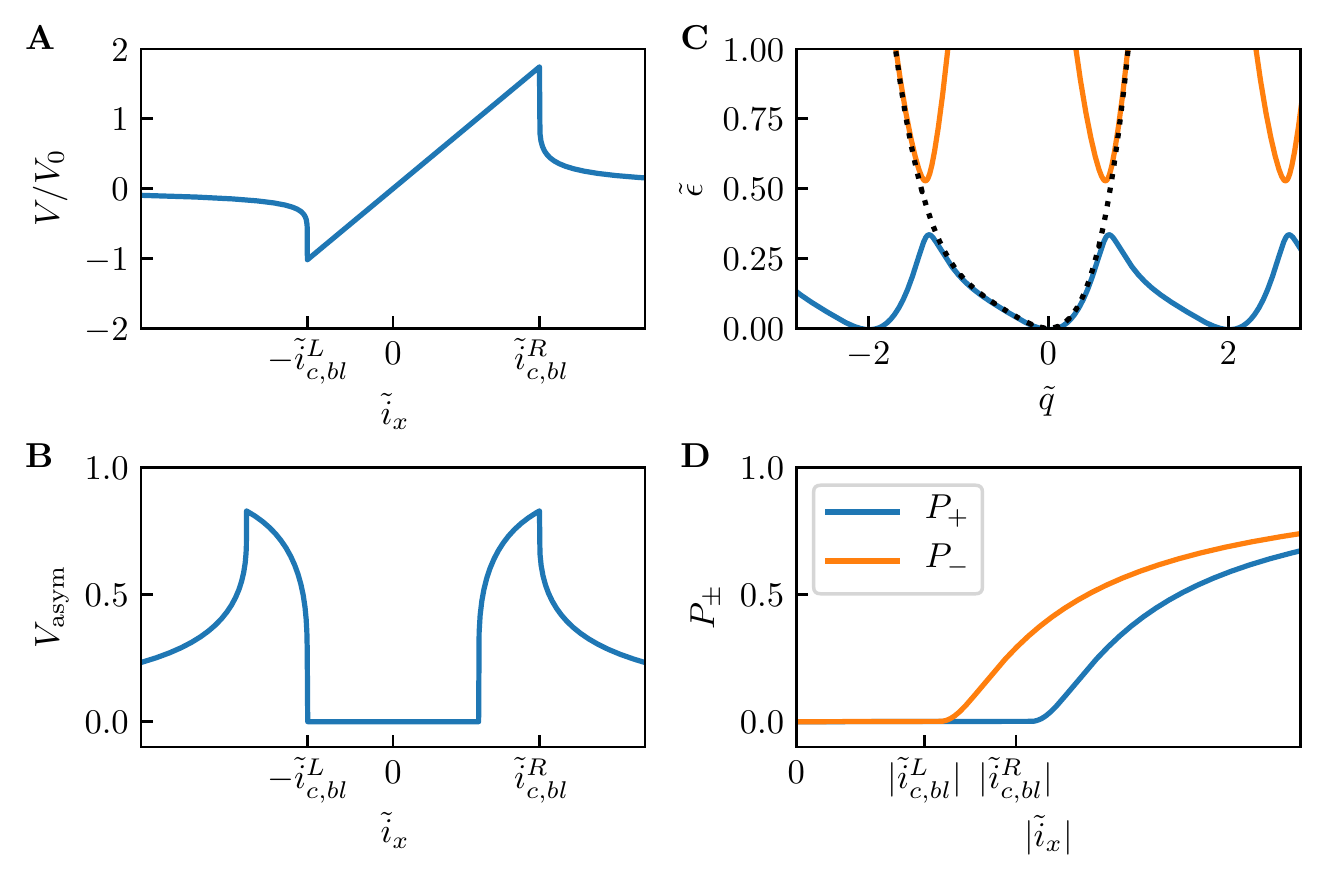}
    \caption{\textbf{$I_x-V$ curve, energy dispersion and nonreciprocal Zener tunneling for case (II).} (\textbf{A}) $I_x-V$ curve and (\textbf{B}) $V_{\rm asym}(\tilde{i}_x) = [V(\tilde{i}_x)+V(-\tilde{i}_x)]/[V(\tilde{i}_x)-V(-\tilde{i}_x)]$ in the presence of the Bloch oscillation, calculated from Eqs. \eqref{eq:bloch_time} and \eqref{eq:bloch_voltagedrop}, where $V_0 = e/C$. (\textbf{C}) Energy dispersion of the two lowest energy bands with the asymmetric changing energy $E'_{ch}(\tilde{q}) = \tilde{q}^2/2 + \tilde{A}\tilde{q}^3+\tilde{A}'\tilde{q}^4$ with $\tilde{A}=0.6$ and $\tilde{A}'=0.3$, and we set $E_J/E_Q=0.2$, where $E_Q=e^2/(2C)$, to open up a gap in the spectrum. Dotted curve represents the energy dispersion without the Josephson coupling term $E_J\cos\varphi$ in the Hamiltonian.
    (\textbf{D}) The LZ rate calculated from Eq. \eqref{Zenerexp} with $E_J/E_Q = 0.1$ ($E_Q=e^2/(2 C)$) and $R/R_q=100$.}
    \label{fig:energy_dispersion}
\end{figure*}
\subsection{Nonreciprocal Bloch oscillation for case (II)}
First, we will discuss the effect of nonreciprocity in Bloch oscillation in Josephson junction. For the energy dispersion \eqref{eq:blochenergy}, denoting the left and right Brillouin zone boundary $\tilde{q}_{L,R}$, the conditions for the Bloch oscillation for $\tilde{i}_x>0$ and $\tilde{i}_x <0$ cases can be written as,
\begin{equation}
    \tilde{i}_x \geq \frac{\partial \tilde{\epsilon}(\tilde{q}_{R})}{\partial \tilde{q}} \eqqcolon \tilde{i}_{c, bl}^R, \quad
\tilde{i}_x \leq \frac{\partial \tilde{\epsilon}(\tilde{q}_{L})}{\partial \tilde{q}} \eqqcolon -\tilde{i}_{c, bl}^L, \label{eq:def_icbl}
\end{equation}
respectively. The periods of the Bloch oscillation for $\tilde{i}_x >0$ and $\tilde{i}_x <0$ cases are,
\begin{equation}
    \tilde{\tau}_{R} = \int_{\tilde{q}_L}^{\tilde{q}_R} \frac{d\tilde{q}}{\tilde{i}_x - \frac{\partial \tilde{\epsilon}}{\partial \tilde{q}}}, \quad
    \tilde{\tau}_{L} = \int_{\tilde{q}_R}^{\tilde{q}_L} \frac{d\tilde{q}}{\tilde{i}_x - \frac{\partial \tilde{\epsilon}}{\partial \tilde{q}}}. \label{eq:bloch_time}
\end{equation}
The voltage drop can be calculated by Eq. \eqref{eq:blochequation} as \cite{likharev_theory_1985}
\begin{equation}
    V_{L,R} = \frac{e}{C} \left< \frac{\partial \tilde{\epsilon}}{\partial \tilde{q}} \right>
    = \frac{e}{C}\left(\tilde{i}_x - \frac{2}{\tilde{\tau}_{L,R}} \right). \label{eq:bloch_voltagedrop}
\end{equation}
We show the voltage drop calculated by Eq. \eqref{eq:bloch_voltagedrop} in Figs. \ref{fig:energy_dispersion}A and B. As we can see, since the critical currents where the Bloch oscillation sets in are different for $\tilde{i}_x>0$ and $\tilde{i}_x<0$, i.e., $\tilde{i}_{c, bl}^R\neq \tilde{i}_{c, bl}^L$, $I_x-V$ curve exhibits nonreciprocity.

\subsection{Nonreciprocal Zener tunneling for case (II)}
Next, we discuss the nonreciprocity in Zener tunneling rate. The general expression of the Zener tunneling rate was derived in Ref. \onlinecite{zaikin_effect_1992}, where the argument is only for the quadratic charging energy. Generalizing their argument to include the asymmetry of the charging energy, we obtain
\begin{align}
    P_{\pm} =
    \exp\left[-\left(\frac{\pi E_J}{2 E_Q}\right)^2\frac{R}{R_q}\frac{1}{|V_{C, \pm}||v_{\pm}|}\right],\quad \left(R_q = \frac{e^2}{2\pi\hbar}\right) \label{Zenerexp}
\end{align} 
where we neglected the effect of the fluctuation of the charge. Here 
\begin{align*}
    V_{C,\pm} = \left.\frac{d}{d\tilde{q}}(E'_{\rm ch}(\tilde{q}) -E'_{\rm ch}(\tilde{q}\mp 2))\right|_{\tilde{q}=\tilde{q}_{R}/\tilde{q}_L},
\end{align*}
and, as we can easily see, $|V_{C,+}|=|V_{C,-}|$. $v_{\pm}$ is the velocity of the charge at $\tilde{q}_{R,L}$ given by the solution of Eq. \eqref{eq:blochequation}, i.e.,
\begin{align}
    v_{\pm} = \tilde{i}_x - 
    \left.\frac{\partial \tilde{\epsilon}}{\partial \tilde{q}}\right|_{\tilde{q}=\tilde{q}_R/\tilde{q}_L}
    = \tilde{i}_x \mp \tilde{i}_{c,bl}^{R/L},
    \label{vpmeq}
\end{align}
where $\tilde{i}_{c,bl}^{R/L}$ are defined in Eq. \eqref{eq:def_icbl}. As we noted $\tilde{i}_{c,bl}^{R} \neq \tilde{i}_{c,bl}^{L}$, so $|v_{+}(\tilde{i}_x)|\neq |v_{-}(-\tilde{i}_x)|$ and $P_{+}\neq P_{-}$. The Landau-Zener tunneling probability $P_{\pm}$ obtained from Eq. \eqref{Zenerexp} is shown in Fig. \ref{fig:energy_dispersion}D. We can see the threshold behavior coming from the dissipation \cite{golubev_effect_1990}.

Here we note the importance of the effect of dissipation in obtaining the nonreciprocal Zener tunneling rate. In the present semiclassical approximation, there occurs no quantum tunneling when the classical solution does not reach the band crossing point due to the dissipation. Then the asymmetric threshold current is the origin of the nonreciprocal tunneling rate, and hence the dissipation is required for the nonreciprocity. On the other hand, it was shown in Ref. \onlinecite{kitamura_nonreciprocal_2019} that the nonreciprocal Landau-Zener tunneling occurs if we have nonzero shift vector even without the dissipation. Here, as we will discuss in Materials and Methods, we are considering the system where $\mathcal{P}'$ and $\mathcal{T}'$ is broken by the asymmetry of the dispersion relation, but the system still has $\mathcal{P'T'}$ symmetry. Then, from the general transformation rule \cite{morimoto_topological_2016}, the shift vector is identically zero. Furthermore, we can show that, in the absence of the shift vector, there is no nonreciprocity in the LZ rate even in the presence of the asymmetry in the band energy. To show this, we observe that, in the absence of shift vector, the amplitude for the tunneling process during one cycle of Bloch oscillation under the electric field $E=-E_x<0$ is given as \cite{kitamura_nonreciprocal_2019},
\begin{align*}
    a_+^{(-E_x)} = i e^{i\arg A_{+-}(-\pi)}
    \int_{-\pi}^{\pi}
    dk_1 |A_{+-}|(k_1)e^{ -i\int_{-\pi }^{k_1}dk_2 \frac{\Delta(k_2)}{-e E_x} }, \numberthis \label{LZ1}
\end{align*}
where $A_{+-}=\braket{u_+|\partial_k|u_-}$, $\ket{u_{\pm}}$ is the wavefunction for upper/lower band, and $\Delta(k)$ is the $k$ dependent difference of the upper band energy and the lower band energy. Although the standard estimation utilizes the integration path in the complex $k$ plane, here we only consider the integration path on the real $k$ line. From Cauchy's theorem, this does not spoil any generality of our result. Then, the expression for the reverse process with the electric field $E=E_x>0$ is given as,
\begin{align*}
    a_+^{(E_x)} = i e^{i\arg A_{+-}(\pi)}
    \int_{\pi}^{-\pi}
    dk_1 |A_{+-}|(k_1)e^{ -i\int_{\pi }^{k_1}dk_2 \frac{\Delta(k_2)}{eE_x} },
\end{align*}
By taking the complex conjugate of Eq. \eqref{LZ1}, we can show that $\left(a_+^{(-E_x)}\right)^* = e^{i\chi}a_+^{(E_x)}$, where 
\begin{align*}
    \chi = -\arg A_{+-}(-\pi) - \arg A_{+-}(\pi) - \int_{-\pi }^{\pi}dk_2 \frac{\Delta(k_2)}{eE_x}.
\end{align*}
Therefore, we conclude that $\left|a_+^{(-E_x)}\right|=\left|a_+^{(E_x)}\right|$ in the absence of shift vector, even if the system breaks $\mathcal{P}$ symmetry. The situation is different if we include the dissipation to the system, as we can see from Eq. \eqref{Zenerexp}. Since the semiclassical dynamics of $Q$ reflects the asymmetry of the dispersion through the dissipative term, the nonreciprocal LZ effect is realized.

\section{Discussion}
\subsection{Nonlinear capacitance}
Here, we estimate the nonlinear capacitance $\alpha$ \cite{buttiker_mesoscopic_1993,buttiker_capacitance_1993,christen_low_1996,ma_weakly_1999,wang_nonlinear_1999} using the scaling form derived by the Thomas-Fermi approximation \cite{ma_weakly_1999,wang_nonlinear_1999}:
\begin{align}
    \alpha &\propto \left[\frac{(4\pi)^2}{\epsilon_{F,2}}
    \left(S\lambda_2 \lambda_2^{-2}e^{-2}\right)^{-2}-
    \frac{(4\pi)^2}{\epsilon_{F,1}}
    \left(S\lambda_1 \lambda_1^{-2}e^{-2}\right)^{-2}
    \right]\frac{1}{e^3},\nonumber\\
    &\propto \left[\frac{1}{n_2}
    -\frac{1}{n_1}
    \right]\frac{4\pi}{e S^2}
\end{align}
where $S$ is the area of the cross section of the Josephson Junction, $\lambda_{1/2}$, $\epsilon_{F,1/2}$ and $n_{1/2}$ are the Thomas-Fermi screening lengths, the Fermi energy and the carrier density of the bulk superconductors, and we replaced $d/d\epsilon$ with $1/\epsilon_F$ ($\epsilon_F$ is the Fermi energy) for the order estimation. Now, the linear capacitance in the Thomas-Fermi approximation can be written as
\begin{equation}
    C = \frac{\epsilon_r}{4\pi} \frac{S}{a + \lambda_1 + \lambda_2},
    \label{eq:capacitance_formula}
\end{equation}
where $\epsilon_r$ and $a$ are the relative dielectric constant and the thickness of the thin film, respectively. 

First we consider the case (I), where the dynamics is governed by Eqs. \eqref{rsjeq2} and \eqref{asymchaene}. Then, in the dimensionless unit, we get
\begin{align}
    A &= \alpha C^{3/2}\sqrt{E_J}\nonumber\\
    &\propto \left[\frac{\epsilon_r }{n_2S(a + \lambda_1 + \lambda_2)}
    -\frac{\epsilon_r }{n_1S(a + \lambda_1 + \lambda_2)}
    \right] \sqrt{\frac{E_J}{2E_Q}}
\end{align}
Now, we set the typical values $n_{1,2} \sim 10^{20}\,{\rm cm^{-3}}$, $\epsilon_r\sim 10$, $S \sim 0.1\,{\rm \mu m^{2}}$, $a = 1 \,{\rm nm}$, $E_Q/E_J \sim 10^{-1}$ and assume $a \gg \lambda_{1,2}$. Then, $A \sim 10^{-3}$.

If we consider the case (II), where the dynamics is governed by Eqs. \eqref{eq:blochequation} and \eqref{eq:blochenergy}, in dimensionless unit,
\begin{align}
    \tilde{A} = \alpha C e \sim \left[\frac{\epsilon_r }{n_2S(a + \lambda_1 + \lambda_2)}
    -\frac{\epsilon_r }{n_1S(a + \lambda_1 + \lambda_2)}
    \right].
\end{align}
Since $E_J \propto S$ and $E_Q \propto 1/S$, $E_J/E_Q \ll 1$ is satisfied for the system with small $S$. Therefore, we assume small Josephson junction and set $S = 0.01 {\rm \mu m^{2}}$, $n_{1,2} \sim 10^{20}\,{\rm cm^{-3}}$, $\epsilon_r\sim 10$, $a = 1 \,{\rm nm}$ and assume $a \gg \lambda_{1,2}$. Then, $\tilde{A} \sim 10^{-2}$.

\subsection{Experimental measurement}
From the above estimate, $A \sim 10^{-3}$ for case (I) and $\tilde{A}\sim 10^{-2}$ for case (II), so the asymmetry is relatively small in the experimental settings, but it is possible to measure the $2\omega$ response $V_{2\omega}$ to the AC driving current $I_{x}(t)=I_a \cos\omega t$ with small $\omega$ with a high precision. Assuming $\omega$ is small compared to the characteristic frequency of the dynamics, we can calculate the $2\omega$ component of the response voltage by the adiabatic approximation:
\begin{align}
    V_{2\omega} &= \frac{\omega}{2\pi}\int_0^{2\pi/\omega} dt \cos(2\omega t) V(I_a\cos\omega t)\nonumber \\
    &= \frac{1}{4\pi}\int_0^{2\pi} d\tau \cos\tau \left[V\left(I_a\cos\frac{\tau}{2}\right)+V\left(-I_a\cos\frac{\tau}{2}\right)\right].
\end{align}

Now, we estimate $V_{2\omega}$ for three cases: (A): case (I) with $T=0$, (B): case (I) with $T>0$ and (C): case (II). As we discussed, the asymmetry of $V$ is pronounced near the various critical value of $i_x$ or $\tilde{i}_x$, so, to obtain large $V_{2\omega}$ we set the amplitude of the external voltage $I_a$ near these critical currents, i.e., (A) $I_c$, (B) $i_{c2}I_c$ and (C) $\tilde{i}_{c,bl}^{L/R}e/(RC)$.

For the case (A), i.e., case (I) with $T=0$, if we set $I_0 > I_C$, the above measurement of $2\omega$ component reflects the difference of $i_{c1}^R$ and $i_{c1}^L$. We set the critical current density $I_c/S = 100 \, {\rm A/cm^2}$ and the resistance times area $RS = 10^{-5}\,{\rm \Omega \, cm^{2}}$, and the capacitance $C/S \sim 10^{-5}\,{\rm F/cm^2}$, where we used Eq. \eqref{eq:capacitance_formula} with $a=1\,{\rm nm}$ and $\epsilon_r=10$. Then we get $r^{-1} \sim 0.1$, and for $A\sim 10^{-3}$, $A'=0.5A$, the numerical calculation yields $V_{2\omega} \sim 0.001 R I_c \sim 1 \, {\rm \mu V}$.

Next, we consider the case (B), i.e., case (I) with $T>0$. We use the same parameters as the case (A) and set $T=50 \,{\rm K}$. Then, the numerical calculation yields $V_{2\omega} \sim 0.001 R I_c \sim 1 \, {\rm \mu V}$.

For the case (C), i.e., case (II), for $\tilde{A} = 0.01$ and $\tilde{A}'=0.5\tilde{A}$, the numerical calculation yields $V_{2\omega} \sim 0.01 e/C \sim 1 \, {\rm \mu V}$, where we used the parameters $C/S \sim 10^{-5}\,{\rm F/cm^2}$ and 
$S=0.01 {\rm \mu m^{2}}$.

In summary, $V_{2\omega}$ is about $1\, {\rm \mu V}$ for the usual Josephson junction systems, and it can be measured by the current experimental technology. As concrete superconducting materials, it is better to use different superconductors with the different carrier density in the normal state, so that the nonlinear capacitance becomes large.

\subsection{Conclusion}
We have shown that, in inversion asymmetric Josephson junctions, the nonreciprocal $I_x-V$ curve is realized if we include the asymmetry of the charging energy both for the system with $E_J/E_Q \gg 1$ and $E_J/E_Q \ll 1$. As we discussed above, the nonreciprocity induced by the nonlinear capacitance can be detected in the current experimental technology. 

\section{Materials and methods}
\subsection{Model for case (I)}
The dc Josephson effect is described by the constant $\varphi$, $Q$ and $V=0$, where $\varphi$ is determined by $I_x = \frac{2e E_J}{\hbar} \sin \varphi=I_c \sin \varphi$. For $|I_x| > I_c$, there is no solution for constant $\varphi$ and the voltage $V$ appears. In this picture, $I_c$ is identical for both directions, while one needs to solve the dynamics, i.e., the time dependence, of $Q$ and $\varphi$ when finite voltage appears. In this case, the functional form of  $E_{ch}(Q)$, which is related to the voltage $V$ by $V = \frac{\partial E_{ch}}{\partial Q}$, is important. Often the form $E_{ch}(Q) = Q^2/(2C) - V_g Q$ is taken with $C$ being the capacitance and $V_g$ the gate voltage. This seems to break the symmetry between right and left, i.e., $Q$ and $-Q$, but the shift in the origin of $Q$ recovers that symmetry. Therefore, the essential asymmetry between right and left comes from the higher order terms in $Q$ such as
\begin{equation}
    E_{ch} = \frac{Q^2}{2C} + \alpha Q^3 + \alpha' Q^4. \label{cheq}
\end{equation}
Then we consider the following generalized Josephson equation as
\begin{equation}
    \frac{\hbar}{2e} \dot{\varphi} = \frac{\partial E_{ch}}{\partial Q},\quad
    \dot{Q} = I_x + \tilde{I}(t) - I_c \sin\varphi -\frac{1}{R}\frac{\partial E_{ch}}{\partial Q}, \label{rsjeq}
\end{equation}
where we added the fluctuating current $\tilde{I}$ satisfying $\braket{\tilde{I}(t)\tilde{I}(t')}=2(\beta R)^{-1}\delta(t-t')$, to discuss the finite temperature system.

It is useful to rewrite Eqs. \eqref{cheq} and \eqref{rsjeq} with the dimensionless parameters $\tilde{i}=\tilde{I}/I_c$, 
$i_x=I_x/I_c$, $r^{-1}=R^{-1}\sqrt{\hbar/(2e C I_c)}$, 
$A=\alpha C^{3/2} \sqrt{E_J}$, $A' = \alpha' C^2 E_J$ and $\tilde{T}^{-1}=E_J \beta$. Also, we rescale $t$ and $Q$ as $\tau = t \sqrt{2eI_c/(\hbar C)}$ and $q = \sqrt{2e/(\hbar C I_c)}Q$. Then, Eq. \eqref{rsjeq} becomes
\begin{equation}
    \frac{d\varphi}{d\tau} = \frac{\partial \epsilon_{ch}}{\partial q},\quad
    \frac{dq}{d\tau} = i_x + \tilde{i}(t) - \sin\varphi -r^{-1}\frac{\partial \epsilon_{ch}}{\partial q}, \label{rsjeq2}
\end{equation}
where
\begin{equation}
    \epsilon_{ch} = \frac{q^2}{2} + A q^3 + A' q^4. \label{asymchaene}
\end{equation}
Let us discuss here the analogy of Eq. \eqref{rsjeq2} with the particle motion under the periodic potential. The Josephson phase $\varphi$ corresponds to the position $x$, while the charge transfer $q$ to the momentum $p$. In this particle picture, the potential energy is $-\cos x$ and the kinetic energy is $\epsilon_{ch}(q \to p)$. In this sense, one can define the ``time-reversal symmetry'' $\cal{T}'$ and ``inversion symmetry'' $\cal{P}'$ as
\begin{align}
\cal{T}': &x \to x, p \to -p, \\ \nonumber
\cal{P}': &x \to -x, p \to -p. 
\label{eq:T'}
\end{align}
Then, our system breaks both $\cal{P}'$ and $\cal{T}'$, while it preserves $\cal{P}'\cal{T}'$ except the dissipative term in Eqs. \eqref{rsjeq} and \eqref{rsjeq2}. Namely, the periodic potential is inversion symmetric,
while the kinetic energy is asymmetric with respect to $p$ and $-p$. In the quantum mechanical case, this leads to the asymmetric dispersion $\varepsilon(k) \ne \varepsilon(-k)$. 

We will discuss the nonreciprocity of Eq. \eqref{rsjeq2} with Eq. \eqref{asymchaene} for two cases: First, we discuss the system with no thermal fluctuation, at $T=0$. For $|i_x|>1$, where the bias is so strong that the potential barrier disappears, the dynamics is characterized by the limit cycle in $(\varphi,q)$ space. For $|i_x|<1$ and sufficiently small $r^{-1}$, there coexists the stable fixed point and the limit cycle \cite{hanggi_reaction-rate_1990,strogatz_nonlinear_2018}, which represents the metastable steady state. Secondly, we discuss the system with thermal fluctuation at finite temperature $T>0$, where the phase slip is caused by the thermal fluctuation \cite{ivanchenko_josephson_1969,ambegaokar_voltage_1969}. In both cases, we will show that the asymmetry of the charging energy leads to the nonreciprocity.

Here we note that, since the voltage drop $V$ in the presence of $A$ satisfies $V(A,-i_x)=-V(-A,i_x)$, the nonreciprocity characterized by $V_{\rm asym}=[V(A,i_x)+V(A,-i_x)]/[V(A,i_x)-V(A,-i_x)]$ can be rewritten as $[V(A,i_x)-V(-A,i_x)]/[V(A,i_x)+V(-A,i_x)]$, so we calculate the voltage drop $V(A,i_x)$ for positive $i_x$ and change the sign of $A$. In the main text, we fix the parameters $A=\pm 0.6$ and $A'=0.3$.

\subsection{Model for case (II)}
As we mentioned in the introduction, the dynamics in this case is governed by Eq. \eqref{eq:dqdteq}, and $E_J/E_Q$ characterizes the ratio of the band gap to the bandwidth, see Fig. \ref{fig:energy_dispersion}C.

In this case, because of the periodicity of the Brillouin zone, the system starts to exhibit Bloch oscillation, which affects the $I_x-V$ curve in a substantial way \cite{likharev_theory_1985,schon_quantum_1990}. Physically, the Bloch oscillation in $Q$ space corresponds to the cooper pair tunneling through the Josephson junction \cite{likharev_theory_1985}, and it reduces the current flowing through the resistive shunt, so the voltage drop $V$ is suppressed. The Bloch oscillation is hindered by the Zener tunneling process where the state is excited to higher energy bands, and $I_x-V$ curve is determined by the competition between the Bloch oscillation and the Zener tunneling process \cite{schon_quantum_1990,zaikin_effect_1992,kuzmin_charge_1996}.

For the discussion of Bloch oscillation, for simplicity, we work in the lowest order approximation in $E_J$, i.e., we neglect the gap at Brillouin zone boundary but assume the periodic structure of the energy dispersion, $\tilde{E}_{\text ch}$, i.e.,
\begin{equation}
    \tilde{E}_{\text ch}(Q) = \min_{n \in \mathbb{Z}} E_{\text ch}(Q-2 n e).
\end{equation}
Setting $Q = e \tilde{q}$, $t = RC\tilde{\tau}$, $I_x = \tilde{i}_xe/(RC)$, Eq. \eqref{eq:dqdteq} becomes
\begin{equation}
    \frac{d\tilde{q}}{d\tilde{\tau}} = \tilde{i}_x - \frac{\partial \tilde{\epsilon}}{\partial \tilde{q}},\label{eq:blochequation}
\end{equation}
where
\begin{equation}
    \tilde{\epsilon}(\tilde{q}) = \min_{n \in \mathbb{Z}} E'_{\text ch}(\tilde{q} -2 n ),\quad
    E'_{\text ch}(\tilde{q}) = \frac{\tilde{q}^2}{2} + \tilde{A}\tilde{q}^3+\tilde{A}' \tilde{q}^4, \label{eq:blochenergy}
\end{equation}
where $\tilde{A} = \alpha Ce$, $\tilde{A}'=\alpha' C e^2$. We set $\tilde{A} = 0.6$ and $\tilde{A}' =0.3$ in the main text.

\section{Supplementary Materials}
\noindent
Fig. S1. The Josephson circuit.

\noindent
Fig. S2. The bifurcation of the system with $\dot{Q}|_{\rm cap.} \neq 0$.

\noindent
Fig. S3. The bifurcation of the system with $\dot{Q}|_{\rm cap.} = 0$.

\noindent
Fig. S4. The limit cycle for $i_x >1$.

\noindent
Fig. S5. The homoclinic orbit for $i_x =r^{-1}=0$.

\section{Acknowledgments}
\begin{acknowledgments}
This work was supported by JST CREST Grant (JPMJCR1874 and JPMJCR16F1) and JSPS KAKENHI (JP18H03676, JP26103006 and JP18J21329).
\end{acknowledgments}

%\bibliographystyle{ScienceAdvances}
%\bibliography{references_1119}

\appendix
\setcounter{equation}{0}
\setcounter{figure}{0}
\setcounter{table}{0}
\makeatletter
\renewcommand{\theequation}{S\arabic{equation}}
\renewcommand{\thefigure}{S\arabic{figure}}
\begin{widetext}
\pagebreak

\begin{center}
\large{\textbf{Supplemental material for ``Theory of nonreciprocal Josephson effect''}}
\end{center}
\section{Materials and Methods}
The schematic figure of the Josephson circuit discussed in the main text is shown in Fig. \ref{fig:josephson_circuit}.
\begin{figure}[b]
\centering
\begin{circuitikz}[european] \draw
(0,2) -- (1,2) to[barrier,l=J,i>=$I_2$] (1,4)
  -- (-1,4) to[C,l_=C,i<=$I_1$] (-1,2) -- (0,2)
  (1,1) to[short,i>=$I_x$] (1,2)
  (1,2) -- (3,2) to[R,l=R,i>=$I_3$] (3,4)-- (1,4) to[short,i>=$I_x$] (1,5)
;
\end{circuitikz}
\caption{\textbf{The Josephson circuit.} The Josephson circuit, where $C$, $J$ and $R$ represent the capacitor, the Josephson junction, and the resistive shunt, respectively.}
\label{fig:josephson_circuit}
\end{figure}
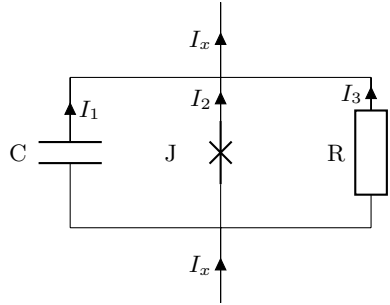

Figs. \ref{fig:bifs_schematic_underdamped} and \ref{fig:bifs_schematic_overdamped} are the details of the bifurcations discussed in the main text. 
To obtain these figures, we used the following generalized Josephson equation
\begin{equation}
    \frac{\hbar}{2e} \dot{\varphi} = \frac{\partial E_{ch}}{\partial Q},\quad
    \dot{Q} = I_x - I_c \sin\varphi -\frac{1}{R}\frac{\partial E_{ch}}{\partial Q}, \label{rsjeq_S}
\end{equation}
where
\begin{equation}
    E_{ch} = \frac{Q^2}{2C} + \alpha Q^3 + \alpha' Q^4. \label{cheq_S}
\end{equation}
We rewrite Eqs. \eqref{rsjeq_S} and \eqref{cheq_S} with the dimensionless parameters
$i_x=I_x/I_c$, $r^{-1}=R^{-1}\sqrt{\hbar/(2e C I_c)}$, 
$A=\alpha C^{3/2} \sqrt{E_J}$, $A' = \alpha' C^2 E_J$. Also, we rescale $t$ and $Q$ as $\tau = t \sqrt{2eI_c/(\hbar C)}$ and $q = \sqrt{2e/(\hbar C I_c)}Q$. Then, Eqs. \eqref{rsjeq_S} and \eqref{cheq_S} becomes
\begin{equation}
    \frac{d\varphi}{d\tau} = \frac{\partial \epsilon_{ch}}{\partial q},\quad
    \frac{dq}{d\tau} = i_x - \sin\varphi -r^{-1}\frac{\partial \epsilon_{ch}}{\partial q}, \label{rsjeq2_S}
\end{equation}
where
\begin{equation}
    \epsilon_{ch} = \frac{q^2}{2} + A q^3 + A' q^4. \label{asymchaene_S}
\end{equation}
Now we discuss the bifurcations for Eqs. \eqref{rsjeq2_S} and \eqref{asymchaene_S} for $r^{-1} \gg 1$ and $r^{-1} \ll 1$.

For $r^{-1} \gg 1$, we can neglect the inertia term (the capacitance) and the equation becomes
\begin{equation}
 r^{-1}d\varphi/d\tau=i_x-\sin\varphi. \label{eq:overdamped_S}
\end{equation}
For $i_x>1$, $d\varphi/d\tau>0$ and there is only a limit cycle (Fig. \ref{fig:bifs_schematic_overdamped}C). At $i_x=i_{c3} = 1$, the saddle-node (blue-sky) bifurcation leads to the vanishing of the limit cycle and the birth of the stable and unstable fixed points at $\varphi=\sin^{-1}i_x$ and $\pi-\sin^{-1}i_x$ for $i_x<1$, respectively, see Figs. \ref{fig:bifs_schematic_overdamped} B and C. For $i_x<1$, the long time dynamics is governed by the stable fixed point, see Figs. \ref{fig:bifs_schematic_overdamped}A and B. Therefore, in this case the disappearance of the limit cycle and the birth of the stable fixed point occur simultaneously, i.e., $i_{c1}^R=i_{c3}=1$. Above $i_{c3}$, the flow of $\varphi$ occurs, and the time-average of $d \varphi/d \tau$ gives that of the voltage drop $\bar{V}={\rm sign}(I_x) R\sqrt{I_x^2 - I_c^2 }$ as we mentioned in the introduction in the main text.

For $r^{-1}\ll 1$, we cannot neglect the inertia term (the capacitance) and the bifurcation mentioned above splits into two bifurcations. One is at $i_x=i_{c3} = 1$, where the saddle-node bifurcation leads to the birth of the stable fixed point and the saddle point at $(\varphi,q)=(\sin^{-1}i_x,0)$ and $(\pi-\sin^{-1}i_x,0)$, as is shown in Figs. \ref{fig:bifs_schematic_underdamped} C, D and E; The other one is the homoclinic bifurcation at $i_x=i_{c1}^R$, where the limit cycle collides with the saddle point at $(\varphi,q)=(\pi-\sin^{-1}i_x,0)$ to become the homoclinic orbit and then disappears, as is shown in Figs. \ref{fig:bifs_schematic_underdamped}A and B. As for the bifurcations for $i_x < 0$, the qualitative nature of the bifurcations are the same, but importantly, $i_{c1}^L \neq i_{c1}^R$ because of the asymmetry of the charging energy.

Figs. \ref{stablemfd}A and B are the limit cycle for Eqs. \eqref{rsjeq2_S} and \eqref{asymchaene_S}. We can see that the limit cycle is different for $A>0$ case (panel A) and $A<0$ case (panel B).

Figs. \ref{homcli}A and B are the homoclinic orbit for Eqs. \eqref{rsjeq2_S} and \eqref{asymchaene_S}. In short, a homoclinic orbit is a variant of a limit cycle. However, in contrast to a limit cycle, there exists a fixed point on it, so its time period is infinite, since it takes infinite time to reach and depart from the fixed point. We can see the obvious difference of the homoclinic orbit for $A>0$ case (panel A) and $A<0$ case (panel B).

\begin{figure*}
    \centering
    \includegraphics[width=\columnwidth]{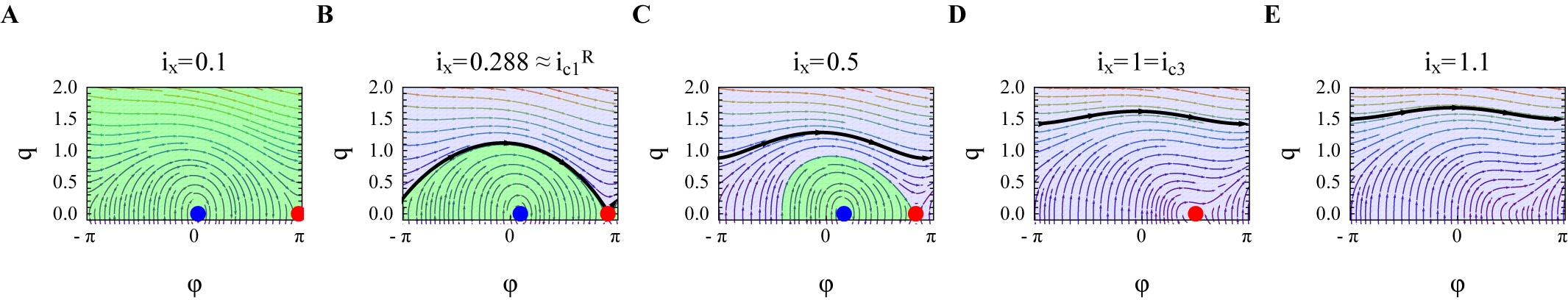}
    \caption{
    \textbf{The bifurcation of the system with} 
    $\mathbf{\dot{Q}|_{\rm cap.}\neq 0.}$
    The bifurcation of the system with finite capacitance,
    Eqs. \eqref{rsjeq2} and \eqref{asymchaene}. We set $A=0.6$, $A'=0.3$, $r^{-1} = 0.1$ and 
    (\textbf{A}) $i_x =0.1$, (\textbf{B}) $i_x =0.288\cong i_{c1}^R$, (\textbf{C}) $i_x =0.5$, (\textbf{D}) $i_x =1=i_{c3}$ and (\textbf{E}) $i_x =1.1$. The blue and red dots represent the stable fixed point and the saddle point, respectively. Black curves are (meta)stable limit cycle, and the green and dark blue regions are the basins of attraction of the stable fixed point (blue dot) and the limit cycle (black curve), respectively. We present the case of positive $i_x$, while the behavior is similar also for $i_x < 0$. However, the critical $i_{c1}^L$ is different from $i_{c1}^R$. }
    \label{fig:bifs_schematic_underdamped}
\end{figure*}
\begin{figure}
    \centering
    \includegraphics[width=0.5\columnwidth]{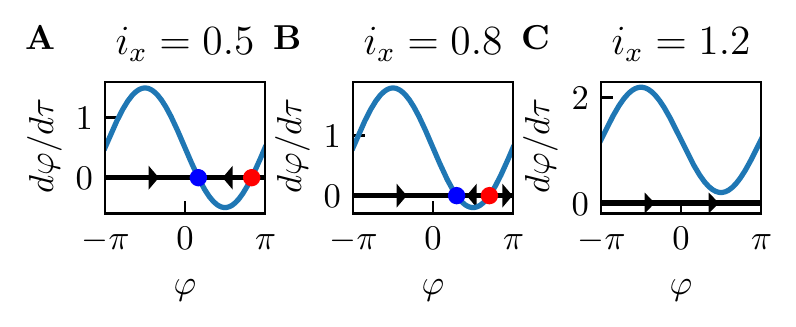}
    \caption{\textbf{The bifurcation of the system with} $\mathbf{\dot{Q}|_{\rm cap.}= 0.}$ The bifurcation of the system with $\dot{Q}|_{cap}=0$, Eq. \eqref{eq:overdamped} for (\textbf{A}) $i_x=0.5$, (\textbf{B}) $i_x=0.8$ and (\textbf{C}) $i_x=1.2$. The blue curve represents $d\varphi/d\tau$ and the arrow on the black curve represents the direction of the velocity. The blue and red dots represent the stable fixed point and the saddle point, respectively. We can see that the limit cycle disappears for $i_x < 1$.}
    \label{fig:bifs_schematic_overdamped}
\end{figure}
\begin{figure}
    \centering
    \includegraphics[width=0.5\columnwidth]{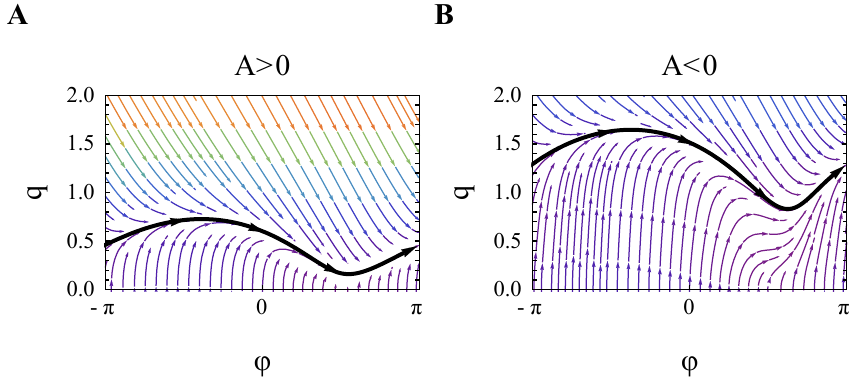}
    \caption{\textbf{The limit cycle for} $\mathbf{i_x>1.}$ The limit cycle, shown by black curves with Eqs. \eqref{rsjeq2} and \eqref{asymchaene} for $i_x = 1.2$, $r^{-1}=1$, and (\textbf{A}) $A>0$ and (\textbf{B}) $A<0$.}
    \label{stablemfd}
\end{figure}
\begin{figure}
    \centering
    \includegraphics[width=0.5\columnwidth]{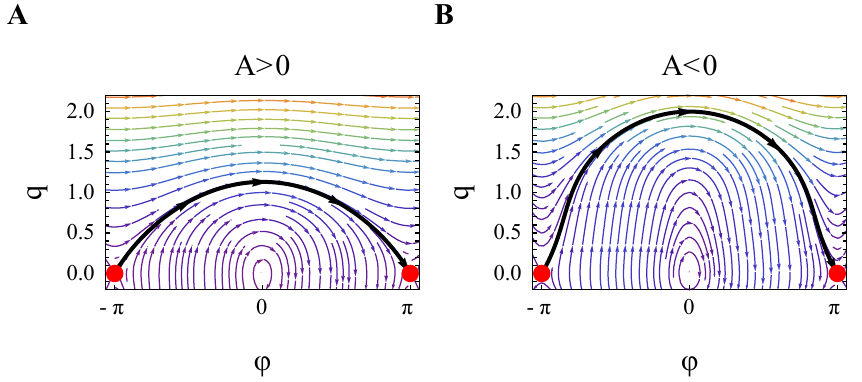}
    \caption{\textbf{The homoclinic orbit for $\mathbf{i_x=r^{-1}=0.}$} The homoclinic orbit (black curves) at $r^{-1}=i_x=0$ for (\textbf{A}) $A>0$ and (\textbf{B}) $A<0$ in Eqs. \eqref{rsjeq2} and \eqref{asymchaene}. The red dots represent the fixed point. Note that $(\varphi,q)=(\pi,0)$ and $(-\pi,0)$ are equivalent.}
    \label{homcli}
\end{figure}
\end{widetext}

\end{document}